\documentclass[11pt]{article}
\usepackage{a4p,cite}
\newcommand{\rb}{\mbox{$R_{\rm b}$}}

\newcommand{\zb}{\mbox{$\rm Z^0$}}
\newcommand{\ccbar}{\mbox{$\rm c\overline{c}$}}
\newcommand{\bbbar}{\mbox{$\rm b\overline{b}$}}

\newcommand{\vcb}{\mbox{$V_{\rm cb}$}}
\newcommand{\mvcb}{\mbox{$|V_{\rm cb}|$}}
\newcommand{\fvcb}{\mbox{${\cal F}(1)|V_{\rm cb}|$}}
\newcommand{\rhsq}{\mbox{$\rho^2$}}
\newcommand{\fw}{\mbox{${\cal F}(\omega)$}}
\newcommand{\fone}{\mbox{${\cal F}(1)$}}
\newcommand{\mean}[1]{\langle{#1}\rangle}
\newcommand{\meanxe}{\mbox{$\mean{x_E}$}}
\newcommand{\bratio}[2]{\mbox{${\rm Br}(#1\rightarrow #2)$}}
\newcommand{\delm}{\mbox{$\Delta m$}}

\newcommand{\wt}{\mbox{$\omega'$}}
\newcommand{\etake}{\mbox{$\eta_{\rm ke}$}}
\newcommand{\tauh}{\mbox{$\hat{\tau_1}$}}
\newcommand{\zetah}{\mbox{$\hat{\zeta_1}$}}
\newcommand{\bplus}{\mbox{$\rm B^+$}}

\newcommand{\bzero}{\mbox{$\rm B^0$}}
\newcommand{\bzerobar}{\mbox{$\rm\bar{B}^0$}}
\newcommand{\bs}{\mbox{$\rm B_s$}}
\newcommand{\bsbar}{\mbox{$\rm \bar{B}_s$}}

\newcommand{\dstar}{\mbox{$\rm D^{*+}$}}
\newcommand{\ddstar}{\mbox{$\rm D^{**}$}}
\newcommand{\dzero}{\mbox{$\rm D^0$}}
\newcommand{\bztodslv}{\mbox{$\bzerobar\rightarrow\dstar\ell^-\bar{\nu}$}}
\newcommand{\brbtodslv}{\mbox{$\rm Br(\bzerobar\rightarrow D^{*+}\ell^-\bar{\nu})$}}
\newcommand{\btodshlv}{$\rm \bar{B}\rightarrow\dstar h\,\ell^-\bar{\nu}$}
\newcommand{\btodsslv}{$\rm \bar{B}\rightarrow\ddstar\ell^-\bar{\nu}$}
\newcommand{\fbd}{\mbox{$f_{\rm B^0}$}}
\newcommand{\taubz}{\mbox{$\rm\tau_{B^0}$}}
\newcommand{\ebzero}{\mbox{$E_{\rm B^0}$}}
\newcommand{\edstar}{\mbox{$E_{\rm D^*}$}}
\newcommand{\elept}{\mbox{$E_{\ell}$}}

\newcommand{\pvbzero}{\mbox{$\bf p_{\rm B^0}$}}
\newcommand{\pvlept}{\mbox{$\bf p_{\ell}$}}
\newcommand{\pvdstar}{\mbox{$\bf p_{\rm D^*}$}}
\newcommand{\pvdzero}{\mbox{$\bf p_{\rm D^0}$}}
\newcommand{\pvneut}{\mbox{$\bf p_{\bar\nu}$}}
\newcommand{\pvvis}{\mbox{$\bf p_{\rm vis}$}}
\newcommand{\qvtx}{\mbox{$Q_{\rm vtx}$}}
\newcommand{\eqvtx}{\mbox{$\sigma_{Q_{\rm vtx}}$}}
\newcommand{\vqvtx}{\mbox{$\sigma^2_{Q_{\rm vtx}}$}}
\newcommand{\PLB}[3] {Phys.~Lett.\ {B#1} (#2) #3}
\newcommand{\PRL}[3] {Phys.~Rev.\ {Lett.~#1} (#2) #3}
\newcommand{\PRD}[3] {Phys.~Rev.\ {D#1} (#2) #3}
\newcommand{\PRP}[3] {Phys.~Rep.\ {#1} (#2) #3}
\newcommand{\NIM}[3] {Nucl.~Instrum.\ {Methods~#1} (#2) #3}
\newcommand{\NPB}[3] {Nucl.~Phys.\ {B#1} (#2) #3}
\newcommand{\CPC}[3] {Comp.~Phys.\ {Comm.~#1} (#2) #3}
\newcommand{\ZPC}[3] {Z.~Phys.\ {C#1} (#2) #3}
\newcommand{\JPH}[3] {J.~Phys.\ {#1} (#2) #3}
\newcommand{\EPJ}[3] {Eur.~Phys.\ J.\ {C#1} (#2) #3}
\newcommand{\IJA}[3] {Int.~J.~Mod.~Phys.\ A.\ {C#1} (#2) #3}
\newcommand{\SJN}[3] {Sov.~J.~Nucl.~Phys.\ {#1} (#2) #3}
\newcommand{\etal} {et~al.}
\input epsf
\newcommand{\epostfig}[3]{
\begin{figure}[tbp]
\setlength{\epsfxsize}{1.1\hsize}
\hspace*{-0.05\hsize} \epsfbox{#1}
\caption{\label{#2}#3}
\end{figure}
}
\newcommand{\fvcbval}{37.5}
\newcommand{\rhsqval}{1.12}
\newcommand{\bdbrval}{5.92}
\newcommand{\fvcbstat}{1.2}
\newcommand{\rhsqstat}{0.14}
\newcommand{\bdbrstat}{0.27}
\newcommand{\fvcbsyst}{2.5}
\newcommand{\fvcbpsyst}{6.8}
\newcommand{\rhsqsyst}{0.29}

\newcommand{\bdbrpsyst}{11.5}
\newcommand{\fvrhcorl}{0.77}
\newcommand{\fvcbeval}{36.8}
\newcommand{\rhsqeval}{1.31}
\newcommand{\bdbreval}{5.11}
\newcommand{\fvcbestat}{1.6}
\newcommand{\rhsqestat}{0.21}
\newcommand{\bdbrestat}{0.19}
\newcommand{\fvcbesyst}{2.0}
\newcommand{\fvcbpesyst}{5.5}
\newcommand{\rhsqesyst}{0.16}
\newcommand{\rhsqevsyst}{0.157}
\newcommand{\bdbresyst}{0.49}
\newcommand{\bdbrpesyst}{9.5}
\newcommand{\fvcbcval}{37.1}
\newcommand{\fvcbcstat}{1.0}
\newcommand{\fvcbcsyst}{2.0}
\newcommand{\rhsqcval}{1.21}
\newcommand{\rhsqcstat}{0.12}
\newcommand{\rhsqcsyst}{0.20}
\newcommand{\bdbrcval}{5.26}
\newcommand{\bdbrcstat}{0.20}
\newcommand{\bdbrcsyst}{0.46}
\newcommand{\statcorl}{0.90}
\newcommand{\systcorl}{0.54}
\newcommand{\fvcboval}{36.9}
\newcommand{\rhsqoval}{0.88}
\newcommand{\fvcbostat}{1.2}
\newcommand{\rhsqostat}{0.14}
\newcommand{\fvrhocrl}{0.79}
\newcommand{\foneval}{0.913}
\newcommand{\foneerr}{0.042}
\newcommand{\vcbval}{40.7}
\newcommand{\vcbstat}{1.1}
\newcommand{\vcbsyst}{2.2}
\newcommand{\vcbtheo}{1.6}
\newcommand{\nhad}{3\,117\,544}
\begin{document}
\begin{titlepage}
{\center\Large

EUROPEAN ORGANIZATION FOR NUCLEAR RESEARCH \\

}
\bigskip

{\flushright
OPAL PR\,305 \\
CERN-EP-2000-032 \\
February 23, 2000\\
}
\begin{center}
    \LARGE\bf\boldmath
    Measurement of \mvcb\ using \bztodslv\ decays
\end{center}
\vspace{1cm}
\bigskip

\begin{center}
\Large The OPAL Collaboration \\
\bigskip
\large
\end{center}
\vspace{1cm}

\begin{abstract}
The magnitude of the Cabibbo-Kobayashi-Maskawa matrix element \vcb\ has been
measured using \bztodslv\ decays recorded on the \zb\ peak
using the OPAL detector
at LEP. The $\dstar\rightarrow\dzero\pi^+$ decays were reconstructed both in 
the particular decay modes $\dzero\rightarrow\rm K^-\pi^+$ 
and $\dzero\rightarrow\rm K^-\pi^+\pi^0$ and 
via an inclusive technique. The product of \mvcb\ and the decay form factor 
of the \bztodslv\ transition at zero recoil \fone\
was measured to be $\fvcb=(\fvcbcval\pm\fvcbcstat\pm\fvcbcsyst)\times 10^{-3}$,
where the uncertainties are statistical and systematic respectively.
By using Heavy Quark Effective Theory calculations for \fone, a value of
\[
\mvcb =(\vcbval \pm \vcbstat \pm \vcbsyst \pm \vcbtheo)\times 10^{-3}
\]
was obtained, where the third error is due to theoretical uncertainties
in the value of \fone. The branching ratio \brbtodslv\ was also measured to be
$(\bdbrcval\pm\bdbrcstat\pm\bdbrcsyst)\,\%$.
\end{abstract}

\vspace{1cm}     

\begin{center}
\large
%To be submitted to Eur.\ Phys.\ J.\ C.
Submitted to Physics Letters B.

\vspace{5mm}

%Please send comments to {\tt richard.hawkings@cern.ch} and \\
%{\tt elisabetta.barberio@cern.ch}
%by 18:00 on Friday 18th February.

\end{center}

\end{titlepage}
\begin{center}{\Large        The OPAL Collaboration
}\end{center}\bigskip
\begin{center}{
%begin authorlist PLEASE DO NOT DELETE THIS COMMENT
G.\thinspace Abbiendi$^{  2}$,
K.\thinspace Ackerstaff$^{  8}$,
P.F.\thinspace Akesson$^{  3}$,
G.\thinspace Alexander$^{ 22}$,
J.\thinspace Allison$^{ 16}$,
K.J.\thinspace Anderson$^{  9}$,
S.\thinspace Arcelli$^{ 17}$,
S.\thinspace Asai$^{ 23}$,
S.F.\thinspace Ashby$^{  1}$,
D.\thinspace Axen$^{ 27}$
% ...
%end authorlist PLEASE DO NOT DELETE THIS COMMENT
}\end{center}\bigskip
\bigskip
%begin institutes
$^{  1}$School of Physics and Astronomy, University of Birmingham,
Birmingham B15 2TT, UK
\newline
$^{  2}$Dipartimento di Fisica dell' Universit\`a di Bologna and INFN,
I-40126 Bologna, Italy
\newline
$^{  3}$Physikalisches Institut, Universit\"at Bonn,
D-53115 Bonn, Germany
\newline
% ...
$^{  8}$CERN, European Organisation for Particle Physics,
CH-1211 Geneva 23, Switzerland
\newline
$^{  9}$Enrico Fermi Institute and Department of Physics,
University of Chicago, Chicago IL 60637, USA
\newline
% ...
$^{ 16}$Department of Physics, Schuster Laboratory, The University,
Manchester M13 9PL, UK
\newline
$^{ 17}$Department of Physics, University of Maryland,
College Park, MD 20742, USA
\newline
% ...
$^{ 22}$Department of Physics and Astronomy, Tel Aviv University,
Tel Aviv 69978, Israel
\newline
$^{ 23}$International Centre for Elementary Particle Physics and
Department of Physics, University of Tokyo, Tokyo 113-0033, and
Kobe University, Kobe 657-8501, Japan
\newline
% ...
$^{ 27}$University of British Columbia, Department of Physics,
Vancouver BC V6T 1Z1, Canada
\newline
% ...
%end notes

%
\section{Introduction}

Within the framework of the Standard Model of electroweak interactions, the
elements of the Cabibbo-Kobayashi-Maskawa mixing matrix
are free parameters, constrained only by the requirement that the matrix
be unitary. The values of the matrix elements can only be determined by 
experiment. Heavy Quark Effective
Theory (HQET) provides a means to extract the magnitude of the 
element \vcb\ from particular semileptonic b decays,
with relatively small theoretical uncertainties \cite{neubrev}.

In this paper, the value of \mvcb\ is extracted by studying
the decay rate for the process \bztodslv\ as a
function of the recoil kinematics
of the \dstar\ meson
\footnote{Charge conjugate reactions are always implied, and the symbol
$\ell$ refers to either an electron or muon.}
\cite{vcbth1,neubfm,isgw}.
The decay rate is parameterised as a function
of the variable $\omega$, defined as the scalar 
product of the four-velocities of the 
\dstar\ and  \bzerobar\ mesons. This is related to the square of
the four-momentum transfer from the \bzerobar\ to the 
$\ell^-\bar\nu$ system, $q^2$, by
\begin{equation}\label{e:wqq}
\omega = \frac{m_{\rm D^{*+}}^2+m_{\rm B^0}^2-q^2}{2m_{\rm B^0}
m_{\rm D^{*+}}},
\end{equation}
and ranges from 1, when the \dstar\ is produced at rest in the \bzerobar\ 
rest frame, to about 1.50. Using HQET, the differential partial width 
for this decay is given by 
\begin{equation}\label{e:decayw}
{\frac{{\rm d} \Gamma({\bztodslv} )}{{\rm d} \omega}}=
\frac{1}{\taubz}\,\frac{{\rm d}\brbtodslv}{{\rm d}\omega}=
{\cal K}(\omega){\cal F}^2(\omega)|V_{\rm cb}|^2\ ,
\end{equation}
where ${\cal K}(\omega)$ is a known phase space term
and \fw\ is the hadronic form factor for
this decay\cite{neubfm}.  Although the shape of this 
form factor is not known, its magnitude
at zero recoil, $\omega=1$, can be estimated using HQET.  
In the heavy quark limit ($m_{\rm b}\to \infty$), \fw\ 
coincides with the Isgur-Wise function \cite{isgw} which is
normalised to unity at the point of zero recoil.
Corrections to \fone\ have been calculated to take into account
the effects of finite quark masses and QCD corrections, yielding
the value and theoretical uncertainty 
$\fone=\foneval\pm\foneerr$ \cite{babar}. 
 Since the phase space factor ${\cal K}(w)$ tends to zero
as $\omega\rightarrow 1$, the decay rate vanishes at
$\omega=1$ and the 
 accuracy of the extrapolation relies on achieving a reasonably
constant reconstruction efficiency in the region near $\omega=1$.  

Previous measurements of \mvcb\ have been made using B mesons produced
on the $\Upsilon(4S)$ resonance \cite{upsvcb} and in \zb\ decays 
\cite{alephvcb,delvcb,opalvcb}. These analyses used a linear or constrained
quadratic expansion of \fw\ around $\omega=1$. An improved theoretical 
analysis, based on dispersive bounds and including higher order corrections,
has since become available \cite{clnff}. This results in a
parameterisation for \fw\
in terms of \fone\ and a single unknown parameter \rhsq\ constrained to lie
in the range $-0.14<\rho^2<1.54$, \rhsq\ corresponding to  the slope of 
\fw\ at zero recoil.

The previous OPAL measurement \cite {opalvcb} used the decay chain
$\dstar\rightarrow\dzero\pi^+$, with the \dzero\ meson being
reconstructed in the exclusive decay channels $\dzero\rightarrow\rm K^-\pi^+$
and $\dzero\rightarrow\rm K^-\pi^+\pi^0$. In this paper, a new analysis is
described in which only the $\pi^+$ from the \dstar\ decay is identified,
and no attempt is made to reconstruct the \dzero\ decay exclusively.
This technique, first employed by DELPHI \cite{delvcb,delb0life},
gives a much larger sample of \bztodslv\ decays than the previous measurement,
but also larger background, requiring a rather more complex analysis.
The measurement of \cite{opalvcb} is also updated to use the new
parameterisation of \fw \cite{clnff}, and improved background models and
physics inputs. In both cases, the initial number of \bzero\ mesons
is determined from other measurements of \bzero\ production in \zb\ decays.

The new reconstruction technique is described 
in Section~\ref{s:rec}, the determination of $\omega$ for each event
in Section~\ref{s:wrec}, the fit to extract \fvcb\ and \rhsq\ in
Section~\ref{s:fit} and the systematic errors in Section~\ref{s:syst}. The
updated exclusive measurement is discussed in Section~\ref{s:excl}
and the measurements are combined and conclusions drawn in 
Section~\ref{s:conc}.

\section{Inclusive reconstruction of \bztodslv\ events}\label{s:rec}

The OPAL detector is well described elsewhere \cite{opaldet}.
The data sample used in this analysis consists of about 4 million hadronic 
\zb\ decays collected during the period 1991--1995, at centre-of-mass energies 
in the vicinity of the \zb\ resonance. Corresponding simulated event samples
were generated using JETSET 7.4 \cite{jetset} as described in \cite{opalrb}.

Hadronic \zb\ decays were selected using standard criteria 
\cite{opalrb}. To ensure the event was well contained within the 
acceptance of the detector, the thrust axis 
direction\footnote{A right handed coordinate system is used, with positive $z$ 
along the electron beam direction and $x$ pointing to the centre of the LEP
ring. The polar and azimuthal angles are denoted by $\theta$ and $\phi$.} 
was required to satisfy $|\cos\theta_T|<0.9$.
Charged tracks and electromagnetic calorimeter clusters with no
associated tracks were then combined into jets using a cone
algorithm~\cite{jetcone}, with a cone half angle of 0.65\,rad and
a minimum jet energy of 5\,GeV. The transverse momentum $p_t$ of each
track was defined relative to the axis of the jet containing it, where
the jet axis was calculated including the momentum of the track. A total
of \nhad\ events passed the event selection.

The reconstruction of \bztodslv\ events was then performed by combining
high $p$ and $p_t$ lepton (electron or muon)
candidates with oppositely charged pions
from the $\dstar\rightarrow\dzero\pi^+$ decay. 
Electrons were identified and photon conversions rejected using neural
network algorithms \cite{opalrb}, and muons were identified as in 
\cite{muonid}. Both electrons and muons were required to have 
momenta $p>2\rm\,GeV$,
transverse momenta with respect to the jet axis $p_t>0.7\rm\,GeV$, and
to lie in the polar angle region $|\cos\theta|<0.9$. 

The event sample was further enhanced in semileptonic b decays by requiring a 
separated secondary
vertex with decay length significance $L/\sigma_L>2$ in any jet of the event.
The vertex reconstruction algorithm and decay length significance 
calculation are described fully in \cite{opalrb}. Together with the lepton 
selection, these requirements result in a sample which is about 90\,\% pure 
in \bbbar\ events.

An attempt was made to estimate the \dzero\ direction in each jet containing
a lepton candidate. Each track (apart from the lepton) and 
calorimeter cluster in the jet was assigned a weight corresponding to 
the estimated probability that it came from the
\dzero\ decay. The track weight was calculated from an artificial neural 
network, trained to separate tracks from b decays and fragmentation tracks
in b~jets \cite{opalrb}. The network inputs are the track momentum, transverse
momentum with respect to the jet axis, and impact parameter
significances with respect
to the reconstructed primary and secondary vertices (if existing). The 
cluster weights were calculated using their energies and angles with respect
to the jet axis alone, the energies first being corrected by subtracting
the energy of any charged tracks associated to the cluster \cite{opalmt}.

Beginning with the track or cluster with the largest weight, 
tracks and clusters were then grouped together until the invariant mass
of the group (assigning tracks the pion mass and clusters zero mass) 
exceeded the charm hadron mass, taken to be 1.8\,GeV. If the final invariant
mass exceeded 2.3\,GeV, the jet was rejected, since Monte Carlo studies
showed such high mass \dzero\ candidates were primarily background.
For surviving jets, the momentum \pvdzero\ of
the group was used as an estimate of the \dzero\ direction, giving
RMS angular resolutions of about 45\,mrad in $\phi$ and $\theta$. The \dzero\
energy was calculated as $E_{\rm D^0}=\sqrt{p_{\rm D^0}^2+m_{\rm D^0}^2}$.

The selection of pions from \dstar\ decays relies on the small mass
difference of only 145\,MeV  \cite{pdg98} between the \dstar\ and \dzero, which
means the pions have very little transverse momentum with respect to the
\dzero\ direction.
Each track in the jet (other than the lepton) was considered as a
slow pion candidate, provided it satisfied $0.5<p<2.5$\,GeV
and had a transverse momentum with respect to the \dzero\ direction
of less than 0.3\,GeV. If the pion under consideration
was included in the reconstructed
\dzero, it was removed and the \dzero\ momentum and energy recalculated.
The final selection was made using the reconstructed mass difference \delm\
between \dstar\ and \dzero\ mesons, calculated as
\[
\delm=\sqrt{E^2_{\rm D^{*}}-|\pvdstar|^2}-m_{\rm D^0}\, ,
\]
where the \dstar\ energy is given by $E_{\rm D^{*}}=E_{\rm D^0}+E_\pi$
and momentum by $\pvdstar={\bf p}_{\rm D^0}+{\bf p}_\pi$.

A new secondary vertex was then iteratively 
reconstructed around an initial seed vertex
formed by the intersection of the lepton and slow pion tracks.
Every other track in the jet was added in turn to the seed vertex, and
the vertex refitted. The track resulting in the lowest vertex fit $\chi^2$
was retained in the seed vertex for the next iteration.
The procedure was repeated until no more tracks could be added without
reducing the vertex fit $\chi^2$ probability to less than 1\,\%. 
The decay length $L'$ between the primary vertex and this secondary vertex, 
and the associated error $\sigma_{L'}$, were calculated as in \cite{opalrb}. 
The pion candidate was accepted if the decay length satisfied 
$-0.1{\rm \,cm}<L'<2$\,cm and the
decay length significance satisfied $L'/\sigma_{L'}>-2$. 

The resulting distributions of \delm\ for opposite and same sign lepton-pion
combinations are shown in Figure~\ref{f:dmdmc}(a) and (b).
The predictions of the Monte
Carlo simulation are also shown, broken down into contributions from
signal \bztodslv\ events, `resonant' background containing real leptons and
slow pions from \dstar\ decay, and combinatorial background,  made up of
events with fake slow pions, fake leptons or both.

\epostfig{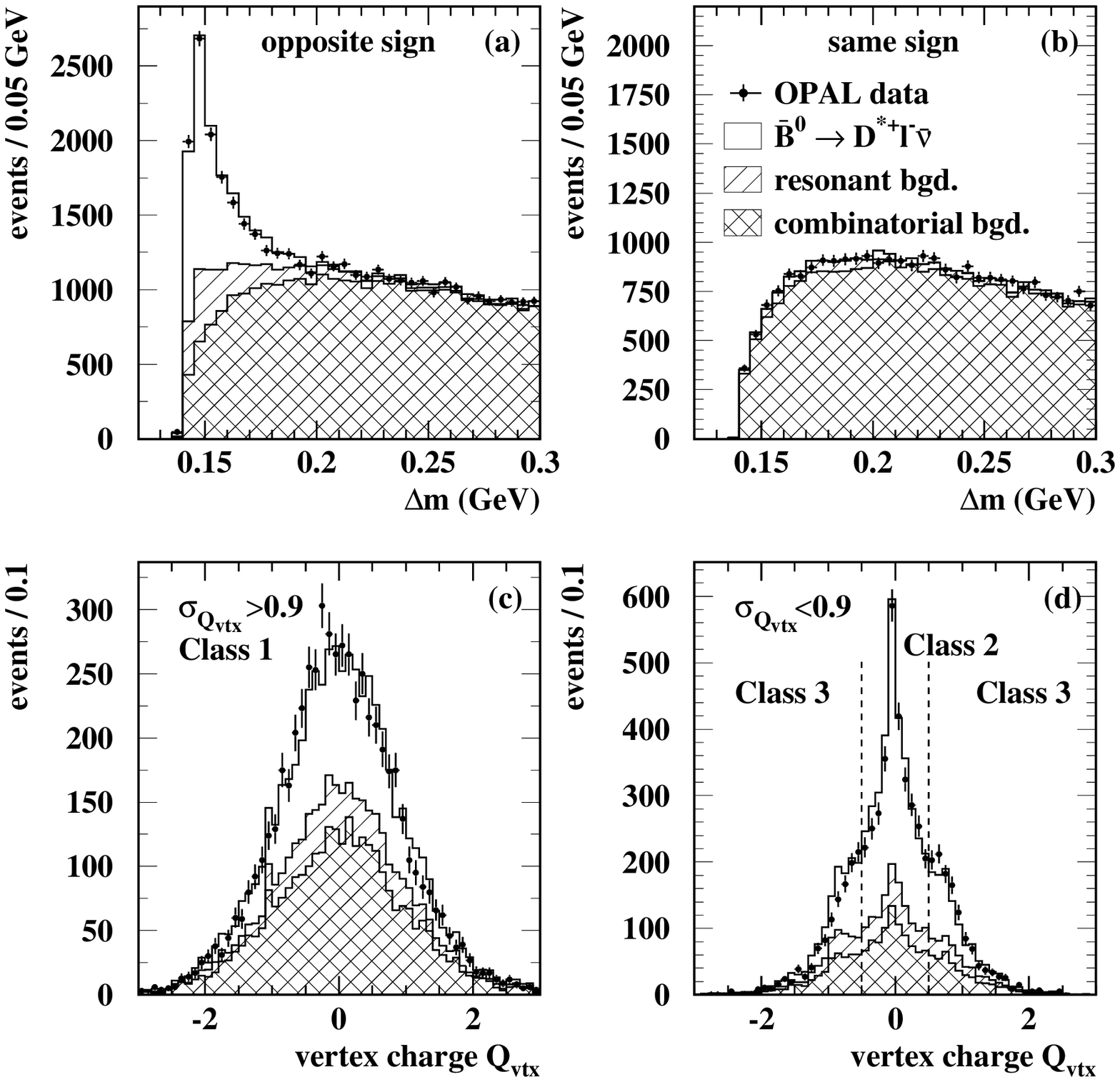}{f:dmdmc}
{Reconstructed \delm\ distributions for selected 
(a) opposite sign and (b) same sign lepton-pion combinations; reconstructed
vertex charge \qvtx\ distributions for opposite sign events with 
$\delm<0.17\,$GeV and (c) $\eqvtx>0.9$ and (d) $\eqvtx<0.9$. In each case
the data are
shown by the points with error bars, and the Monte Carlo simulation
contributions from signal \bztodslv\ decays, other resonant \dstar\ decays 
and combinatorial background are shown by the open, single and cross hatched
histograms respectively. The vertex charge classes 1 (poorly reconstructed),
 2 (well reconstructed neutral vertex) and 3 (well reconstructed charged 
vertex) are indicated.}

In Monte Carlo simulation, about 45\,\% of  opposite sign events with 
$\delm<0.17\rm\,GeV$ are signal 
\bztodslv\ events, 14\,\% are resonant background and
41\,\% are combinatorial background. The 
resonant background is made up mainly of 
$\rm B^-\rightarrow\dstar\pi^-\ell^-\bar\nu$,
$\bzerobar\rightarrow\dstar\pi^0\ell^-\bar\nu$ and
$\bsbar\rightarrow\dstar\rm K^0\ell^-\bar\nu$ decays. These are expected
to be dominated by b semileptonic decays involving orbitally
excited charm mesons (generically referred to as \ddstar),  {\em e.g.\/} 
$\rm B^-\rightarrow D^{**0}\ell^-\bar\nu$ followed by 
$\rm D^{**0}\rightarrow\dstar\pi^-$. These decays will be denoted collectively
by \btodshlv.
Small contributions are also expected
from $\rm b\rightarrow\dstar\tau\bar\nu X$ decays with the $\tau$ decaying 
leptonically, and 
$\rm b\rightarrow \dstar\rm D_s^-X$ with the $\rm D_s^-$
decaying semileptonically (each about 1\,\% of opposite sign events).
For same sign events with $\delm<0.17\,$GeV, there
is a small resonant contribution of about 6\,\% from events with a real 
$\dstar\rightarrow\dzero\pi^+$ where the \dzero\ decays semileptonically,
and the rest is combinatorial background.

The most important background, from \btodshlv\ decays, comes from
both charged \bplus\ and neutral \bzero\ and \bs\ decays, whereas
the signal comes only from \bzero\ decays. Therefore the 
charge \qvtx\ of the reconstructed secondary vertex containing the lepton and 
slow pion and its estimated error \eqvtx\ were calculated, using
\begin{eqnarray*}
\qvtx & = & \sum_i w_i q_i\ , \\
\vqvtx & = & \sum_i w_i (1-w_i) q_i^2 \ ,
\end{eqnarray*}
where $w_i$ is the  weight
for track $i$ of charge $q_i$ to come from the secondary vertex,
and the sums are taken over all tracks in 
the jet \cite{opalbdstar,opaljpks}. The weights were calculated 
in a similar way to those used for the \dzero\ direction reconstruction,
using a neural network with the track momentum, transverse momentum and
impact parameter significances 
with respect to the reconstructed primary and secondary 
vertices as inputs. The weights for the lepton and slow pion candidate tracks
were set to one. 
The vertex charge distributions for opposite sign events with 
$\delm<0.17\,$GeV are shown in Figure~\ref{f:dmdmc}(c) and (d).

The reconstructed vertex charge and error were used to divide the data
into different classes enhanced or depleted in \bplus\ decays,
thus reducing the effect of this background and increasing the statistical
sensitivity.
Three classes $c$ were used---class~1 where the charge is measured poorly
($\eqvtx>0.9$), class~2 where the charge is measured well and is 
compatible with a neutral vertex ($\eqvtx<0.9$, $|\qvtx|<0.5$) and
class~3 where the charge is measured well and is compatible with a 
charged vertex ($\eqvtx<0.9$, $|\qvtx|>0.5$).

\section{Reconstruction of \boldmath$\omega$}\label{s:wrec}

The recoil variable $\omega$ was estimated in each event using the
reconstructed four-momentum transfer to the $\ell\bar{\nu}$ system:
\[
q^2=(\ebzero-\edstar)^2-(\pvbzero-\pvdstar)^2
\] 
together with equation~\ref{e:wqq}. 
The \bzero\ and \dstar\ energies \ebzero\ and 
\edstar\ were estimated directly, whilst the momentum
vectors  \pvbzero\ and \pvdstar\
were estimated using the energies together with the reconstructed polar and
azimuthal angles, as described in more detail below.

Since the slow pion has very little momentum in the rest frame of 
the decaying \dstar, 
the momentum (and hence energy) of the \dstar\ was estimated by scaling the 
reconstructed slow pion momentum by $m_{\rm D^{*+}}/m_{\pi}$,
as for the $\dzero\rightarrow\rm K^-\pi^+\pi^0$ channel
in \cite{opalvcb}. A small correction (never exceeding 12\,\%) was applied,
as a function of $\cos\theta^*$,
the angle of the slow pion in the rest frame of the \dzero.
This procedure gave a fractional \dstar\ energy resolution of 15\,\%.
The polar and azimuthal angles
of the \dstar\ were reconstructed using weighted averages of the
slow pion and \dzero\ directions, giving resolutions of about 22\,mrad on
both $\phi$ and $\theta$.

The energy of the \bzero\ was estimated using a technique similar to that 
described in \cite{opaldil}, exploiting the overall energy and momentum 
conservation in the event to calculate the energy of the unreconstructed
neutrino. First, the energy $E_{\rm bjet}$ of the jet containing the \bzero\ 
was inferred from the measured particles in the rest of the event,
by treating the event as a two-body decay of a \zb\ into
a b jet (whose mass was approximated by the \bzero\ mass)
and another object making up the rest of the 
event. Then, the total fragmentation energy $E_{\rm frag}$ in the b jet
was estimated from the measured visible energy in the b jet and the identified
\bzero\ decay products: $E_{\rm frag}=E_{\rm vis}-\elept-\edstar$. Finally,
the \bzero\ energy was calculated as $\ebzero=E_{\rm bjet}-E_{\rm frag}$,
giving an RMS resolution of 4.4\,GeV.

The b direction was estimated using a combination of two techniques. In the
first, the momentum vector of the \bzero\ was reconstructed from its decay
products:
\[
\pvbzero=\pvdstar+\pvlept+\pvneut,
\]
the neutrino energy being estimated from the reverse of the event visible
momentum vector: $\pvneut=-\pvvis$. The visible momentum
was calculated using all the
reconstructed tracks and clusters in the event, with a correction for 
charged particles measured both in the tracking detectors and calorimeters
\cite{opalmt}. This direction estimate
is strongly degraded if a second neutrino is present ({\em e.g.} from 
another semileptonic b decay in the opposite hemisphere of the event), and its
error was parameterised as a function of the visible energy in the
opposite hemisphere. The resulting resolution is typically between
40 and 100\,mrad for both $\phi$ and $\theta$.

The second method of estimating the \bzero\ direction used the vertex flight
direction---{\em i.e.} the direction of the vector between the primary vertex
and the secondary vertex reconstructed around the lepton and slow pion
as described in Section~\ref{s:rec}. The accuracy of this estimate is 
strongly dependent on the \bzero\ decay length, and was used only if
the decay length significance $L'/\sigma_{L'}$ exceeded 3. After this cut,
the angular resolution varies between about 15 and 100\,mrad, 
and is worse for $\theta$ in the  
1991 and 1992 data where accurate $z$ information from the silicon 
microvertex detector was not available.
The \bztodslv\ candidate was rejected if the two reconstruction methods 
gave $\theta$
or $\phi$ angles disagreeing by more than three standard deviations, which
happened in 7\,\% of Monte Carlo signal events.
Finally, the two \bzero\ direction
estimates were combined according to their estimated 
uncertainties, giving average resolutions of  35\,mrad on $\phi$ and
43\,mrad on $\theta$, including events where only the first method 
was used.

The estimate of $q^2$ derived from the \bzero\ and \dstar\ energies 
and angles was improved by applying the constraint that the mass
$\mu$ of the neutrino produced in the \bztodslv\ decay should be zero.
The neutrino mass is given from the reconstructed quantities by:
\[
\mu^2=(\ebzero-\edstar-\elept)^2-(\pvbzero-\pvdstar-\pvlept)^2\,.
\]
The constraint was implemented by calculating $q^2$ using a kinematic fit,
incorporating the measured values and estimated uncertainties of the 
\bzero\ and \dstar\ energies and angles (the \bzero\ angular uncertainties
varying event by event). This procedure improved the average $q^2$
resolution from 2.78\,GeV$^2$ to 2.57\,GeV$^2$. In Monte Carlo simulation, 
11\,\% of signal events were reconstructed with $q^2<0$, corresponding
to an unphysical value of $\omega$ larger than $\omega_{\max}\approx 1.5$,
and were rejected.

The resulting distributions of reconstructed $\omega$ for various ranges
of true $\omega$ (denoted \wt) in simulated \bztodslv\ decays 
are shown in Figure~\ref{f:resfn}(a--e). The average RMS $\omega$ 
resolution is about 0.12, but there are significant non-Gaussian tails.
The resolution was parameterised (separately for the 1991--2 and 1993--5 data)
as a continuous function $R(\omega,\wt)$ giving the expected distribution 
of reconstructed $\omega$ for each true value \wt. The resolution function
was implemented 
as the sum of two asymmetric Gaussians ({\em i.e.} with different widths
either side of the peak)
whose parameters were allowed to vary as a function of \wt. The 
convolution of this resolution function with the Monte Carlo \wt\
distribution is also shown in Figure~\ref{f:resfn}(a--e), demonstrating 
that the resolution function models the $\omega$ distributions well.

\epostfig{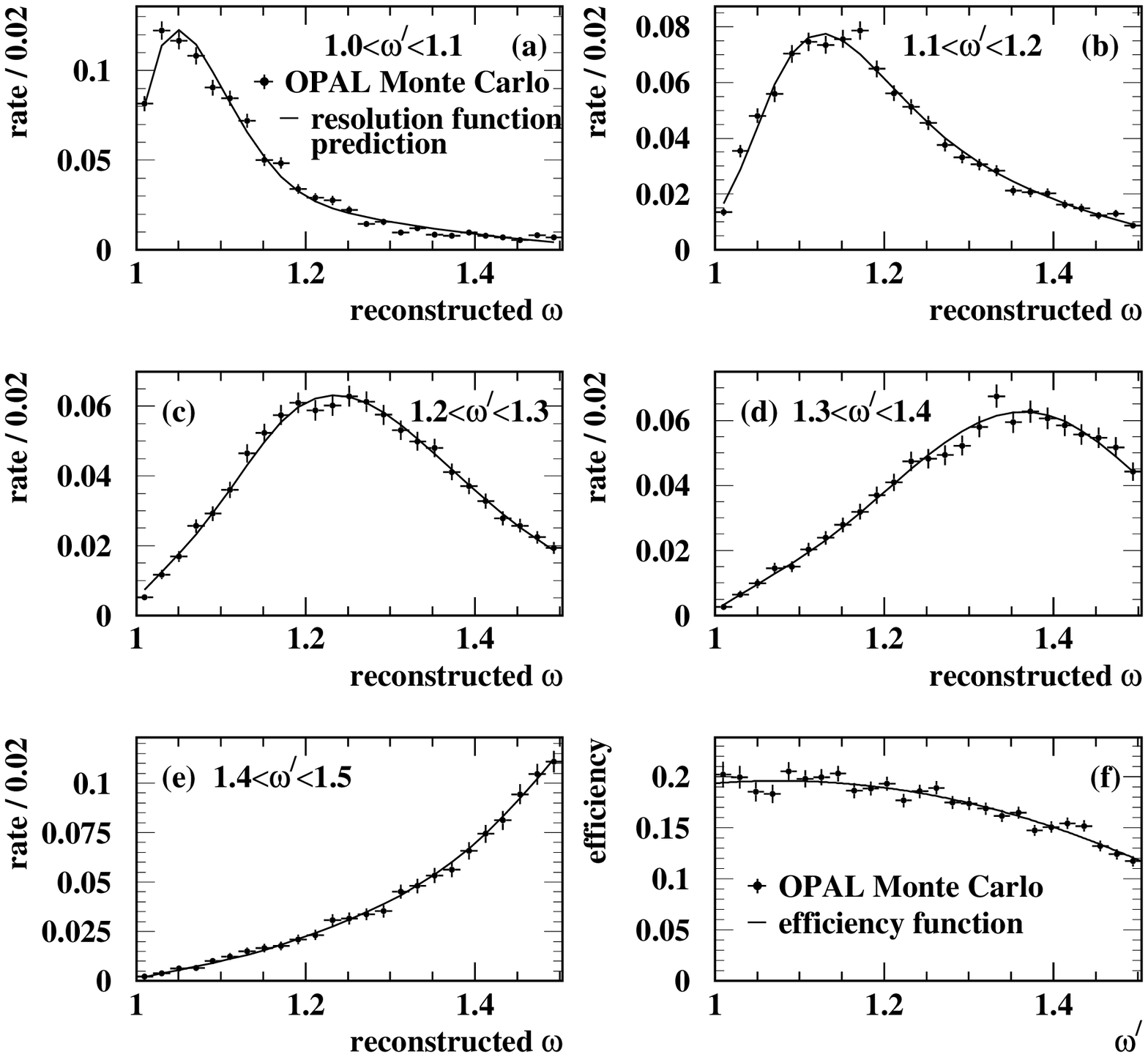}{f:resfn}{Reconstructed $\omega$ distributions for
various ranges of true $\omega$ (denoted \wt) 
in Monte Carlo \bztodslv\ events, together
with the prediction from the resolution function. Events reconstructed
in the unphysical region with $\omega>\omega_{\rm max}$ are rejected.
The reconstruction efficiency as a function of true $\omega$ is also shown.}

The efficiency to reconstruct \bztodslv\ decays $\epsilon(\wt)$ 
is shown in Figure~\ref{f:resfn}(f), together with a second order polynomial
parameterisation. The efficiency varies with \wt, but is reasonably flat
in the critical region near $\wt=1$ where the extrapolation to measure
\fvcb\ is carried out.

\section{Fit and results}\label{s:fit}

The values of \fvcb\ and \rhsq\ were extracted using an
extended maximum likelihood fit to the reconstructed mass difference
\delm,  recoil $\omega$ and vertex charge class $c$ of each event. 
Both opposite and same sign events 
with $\delm<0.3\rm\,GeV$ and $\omega<\omega_{\rm max}$ 
were used in the fit, the high \delm\ and same sign 
events serving to constrain the combinatorial background 
normalisation and shapes in the opposite sign low \delm\ 
region populated by the \bztodslv\ decays. Using the \delm\ value from
each event in the fit, rather than just dividing the data into 
low \delm\ `signal' and high \delm\ 
`sideband' mass regions, increases the statistical sensitivity as the
signal purity varies considerably within the low \delm\ region.

The logarithm of the overall likelihood was given by
\begin{equation}\label{e:totlike}
\ln {\cal L} = \sum_{i=1}^{M^a} \ln {\cal L}^a_i +
\sum_{j=1}^{M^b} \ln {\cal L}^b_j -N^a-N^b
%{\cal L} = e^{-N^r-N^w} \left( \prod_{i=1}^{M^r} {\cal L}^r_i \right) 
%\left( \prod_{j=1}^{M^w} {\cal L}^w_j \right)
\end{equation}
where the sums of individual event log-likelihoods $\ln {\cal L}^a_i$ and 
$\ln {\cal L}^b_j$ are taken over all the observed $M^a$ opposite sign and
$M^b$ same sign events in the data sample, and $N^a$ and $N^b$ are the
corresponding expected numbers of events.

The likelihood for each opposite sign event was given in terms of 
different types or sources of event by
\begin{equation}\label{e:levtr}
{\cal L}^a_i(\delm_i,\omega_i,c_i) 
= \sum_{s=1}^4 N^a_s f^a_{s,c_i} M_{s,c_i}(\delm_i) \, P_s(\omega_i)
\end{equation}
where $N^a_s$ is the number of expected events for source $s$, 
$f^a_{s,c}$ is the fraction of events in source $s$ appearing in 
vertex charge class $c$, $M_{s,c}(\delm)$ is the mass difference distribution
for source $s$ in class $c$ and $P_s(\omega)$ the recoil distribution
for source $s$. For each source, the vertex charge fractions 
$f^a_{s,c}$ sum to one and the mass difference 
$M_{s,c}(\delm_i)$ and recoil $P_s(\omega)$ distributions are normalised to
one. The total number of expected events is given by the sum of the
individual contributions: $N^a=\sum_{s=1}^4 N^a_s$.
%\begin{eqnarray*}
%\sum_s N^r_s & = & N^r \\
%\sum_c f^r_{s,c} & = & 1 \\
%\int^{m_{\rm max}}_{m_\pi} M_{s,c}(\delm)\,d\delm & = & 1 \\
%\int^{\omega_{\rm max}}_1 P_s(\omega)\,d\omega & = & 1 
%\end{eqnarray*}

There are four opposite sign sources: (1) signal \bztodslv\ events, (2)
\btodshlv\ events where the \dstar\ is produced via
an intermediate resonance (\ddstar), (3) other opposite sign background
involving a genuine lepton and a slow pion from \dstar\ decay  and 
(4) combinatorial background. The sum of sources 2 and 3 are shown as
`resonant background' in Figure~\ref{f:dmdmc}.
A similar expression to equation~\ref{e:levtr} was used for ${\cal L}^b_j$,
the event likelihood for same sign events. In this case, only  sources~3
and~4 contribute.

The mass difference distributions $M_{c,s}(\delm)$ 
for sources 1--3 were represented by analytic functions, whose parameters
were determined using large numbers of simulated events, as were
the recoil distributions $P_{s}(\omega)$ for sources~2 and~3. The fractions
in each vertex charge class for sources 1--3 were also taken from simulation.
For the signal (source 1), the product of the expected number of
events $N^a_1$ and recoil distribution $P_1(\omega)$ was given by convolving
the differential partial decay width (equation~\ref{e:decayw}) with the
signal resolution function and reconstruction efficiency:
\[
N^a_1 P_1(\omega)=4N_{\rm Z} R_{\rm b} \fbd \taubz \bratio{\dstar}{\dzero\pi^+}
\int_1^{\omega_{\rm max}} {\cal F}^2(\wt) |\vcb|^2 {\cal K}(\wt) \epsilon(\wt)
R(\omega,\wt)\,d\wt 
\]
where $N_{\rm Z}$ is the number of hadronic \zb\ decays passing the event
selection, $R_{\rm b}\equiv \Gamma_{\rm b\bar{b}}/\Gamma_{\rm had}$ 
is the fraction of hadronic \zb\ decays to \bbbar, 
\fbd\ the fraction of b quarks hadronising to a \bzerobar\ and 
\taubz\ the \bzero\ lifetime. The factor of four accounts for the
two b hadrons produced per $\zb\rightarrow\bbbar$ event and the two
identified lepton species (electrons and muons). 
The form factor ${\cal F}(\wt)$ is given
in \cite{clnff} in terms of the normalisation \fone\ and slope parameter
\rhsq. The efficiency function $\epsilon(\wt)$
and resolution function $R(\omega,\wt)$ were described in 
Section~\ref{s:wrec}, and the known phase space factor 
${\cal K}(\wt)$ is given
in \cite{opalvcb}. The assumed values of the numerical quantities 
are given in Table~\ref{t:input}.

\begin{table}
\centering

\begin{tabular}{l|c|l}
Quantity & Assumed value & Reference \\ \hline
\rb\ & $(21.70\pm 0.09)\,\%$ & \cite{pdg98} \\
\fbd\ & $(39.7^{+1.8}_{-2.2})\,\%$ & \cite{pdg98} \\
\taubz & $1.56\pm 0.04$\,ps & \cite{pdg98} \\
\bratio{\dstar}{\dzero\pi^+} & $(68.3\pm 1.4)\,\%$ & \cite{pdg98} \\
\bratio{\rm b}{\dstar h\,\ell\bar\nu} & $(0.76\pm 0.16)\,\%$ & 
\cite{alephdss}, see text\\
\bratio{\rm b}{\dstar\tau^-\bar\nu X} & $(0.65\pm 0.13)\,\%$ & 
\cite{pdg98}, see text \\
\bratio{\bzerobar}{\dstar\rm D_s^{(*)-}} & $(4.2\pm 1.5)\,\%$ & \cite{pdg98}\\
\bratio{\rm b}{\dstar X} & $(17.3\pm 2.0)\,\%$ & \cite{opaldprod} \\
\bratio{\rm c}{\dstar X} & $(22.2\pm 2.0)\,\%$ & \cite{opaldprod} \\
\bratio{\dzero}{\rm K^-\pi^+} & $(3.85\pm 0.09)\,\%$ & \cite{pdg98} \\
\bratio{\dzero}{\rm K^-\pi^+\pi^0} & $(13.9\pm 0.9)\,\%$ & \cite{pdg98} \\
\hline
\end{tabular}
\caption{\label{t:input} Input quantities used in the fits for
\fvcb\ and \rhsq. The values marked `see text' are derived using methods
explained in Section~\ref{s:fit}.}
\end{table}

Since the data are divided into different vertex charge classes
enhanced and depleted in \bplus\ decays, the fit
gives some information on the amount of \btodshlv\ background.
The predicted level of this background in the fit
was therefore allowed to vary under a Gaussian
constraint corresponding to the branching ratio of 
$\bratio{b}{\dstar\rm h\,\ell\bar\nu}=(0.76\pm 0.16)\,\%$.
The latter has been calculated from the measured branching ratio
$\bratio{\rm b}{\dstar\pi^-\ell\bar\nu X}=(0.473\pm 0.095)\,\%$ 
\cite{alephdss},
assuming isospin and SU(3) flavour symmetry to obtain the corresponding 
$\rm b\rightarrow\dstar\pi^0\ell\bar\nu$ and 
$\rm b\rightarrow\dstar\rm K^0\ell\bar\nu$
branching ratios. A scaling factor of $0.75\pm 0.25$ was included for
the last branching ratio to account for possible SU(3) violation effects
reducing the branching ratio $\rm D_s^{**+}\rightarrow D^{*+}K^0$
compared to the expectation of $\frac{3}{2}\bratio{\rm D^{**+}}{D^{*+}\pi^0}$.
The $P_2(\omega)$ distribution for these events was taken from simulation,
using the calculation of Leibovich~\etal \cite{ligeti} to
predict their recoil spectrum.

The numbers $N_3^{a,b}$ and $P_3(\omega)$ distributions for the small 
background contributions covered by source~3 (both opposite and same sign)
were taken from Monte Carlo simulation, with branching ratios adjusted to the
values given in Table~\ref{t:input}. The branching ratio for
$\rm b\rightarrow\dstar\tau^-\bar\nu$ was derived using the inclusive
branching ratio $\bratio{\rm b}{\tau X}=(2.6\pm 0.4)\,\%$ \cite{pdg98}, 
assuming
a \dstar\ is produced in a fraction $0.25\pm 0.03$ of the time, as seen
for the corresponding decays $\rm b\rightarrow\dstar\ell X$
and $\rm b\rightarrow\ell X$ ($\ell=\rm e,\mu$)
\cite{pdg98}. 
The rate of $\rm b\rightarrow\dstar\ D_s^-X$ was assumed to be dominated by
the two body decay $\bzerobar\rightarrow\dstar\rm D_s^{(*)-}$. The rate
of real \dstar\ decays in the same sign background depends on the
production fractions of \dstar\ in \bbbar\ and \ccbar\ events, which were
taken from \cite{opaldprod}.

The parameters of the analytic functions describing the combinatorial 
background ($N_4^a$, $N_4^b$, $f_{4,c}^a$, $f_{4,c}^b$, $M_{4,c}(\delm)$ and
$P_4(\omega)$) were fitted entirely from the data, with only the choice
of functional forms motivated by simulation. The shapes of the
mass and recoil functions (including a small correlation between
\delm\ and $\omega$) are constrained by the same sign sample (which is
almost entirely combinatorial background), and are the same for each vertex
charge class $c$. The opposite sign high \delm\ region
serves to normalise the number of combinatorial background events
in the low \delm\ region for each vertex charge class.

The values of \fvcb\ and \rhsq\ were extracted by maximising the total
likelihood given by equation~\ref{e:totlike}. The values of
\fvcb\ and \rhsq\ were allowed to vary, together with the level
of \btodshlv\ background and 13 auxiliary 
parameters describing the combinatorial background distributions.
A result of 
\begin{eqnarray*}
\fvcb & = & (\fvcbval \pm \fvcbstat)\times 10^{-3}\,, \\
\rhsq & = & \rhsqval \pm \rhsqstat\
\end{eqnarray*}
was obtained, where the errors are only statistical. The correlation
between \fvcb\ and \rhsq\ is \fvrhcorl. The distributions
of reconstructed $\omega$ for opposite and same sign events with 
$\delm<0.17$\,GeV, together with the fit results, are shown
in Figure~\ref{f:fit1}. The fit describes the data well, both in this region
and the high \delm\ region dominated by combinatorial background, for 
all three of the vertex charge classes.

\epostfig{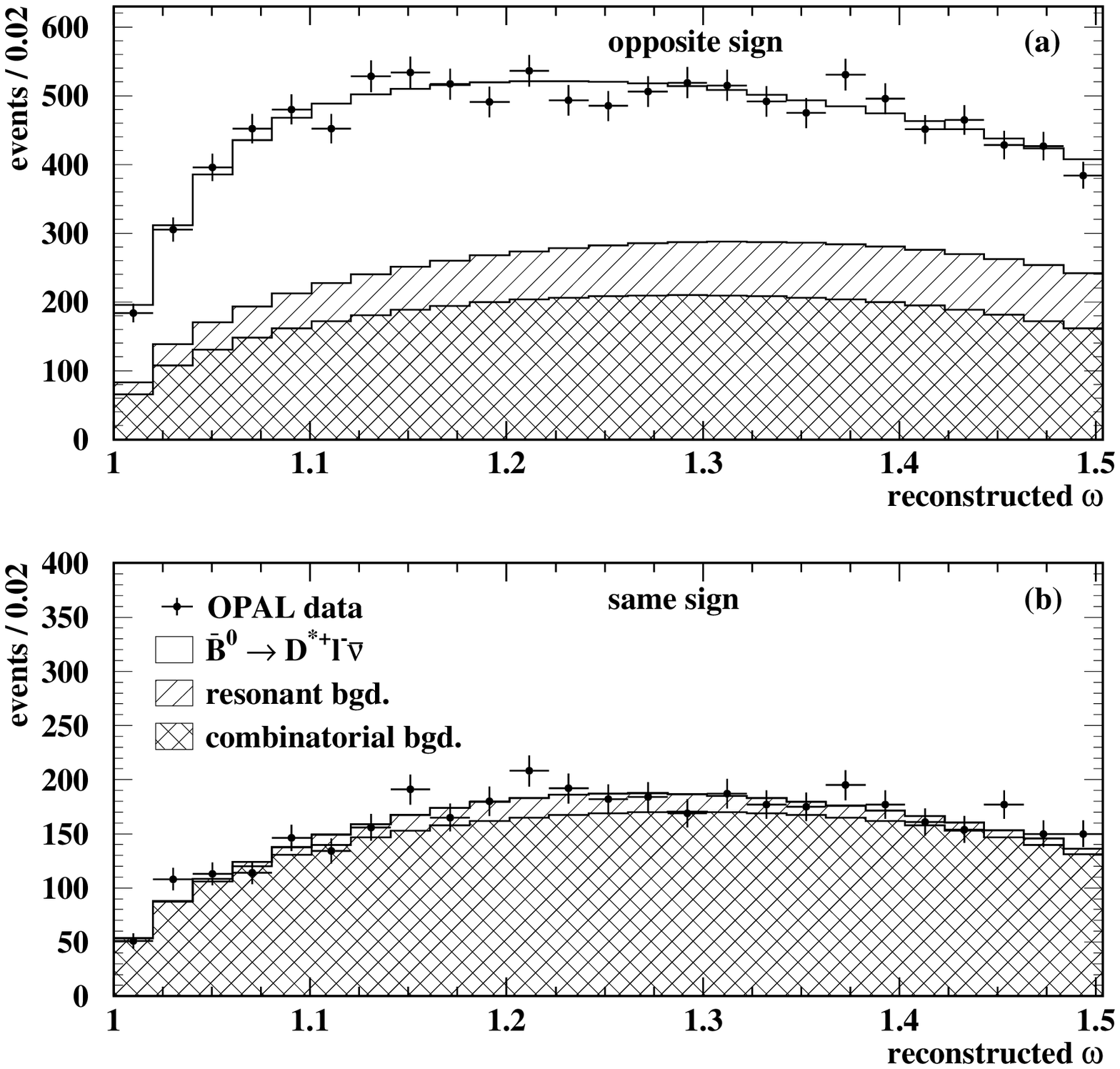}{f:fit1}{Distributions of reconstructed $\omega$ for
(a) opposite sign and (b) same sign events with $\delm<0.17\,$GeV. The
data are shown by the points with error bars and the expectation from the 
fit result by the histograms. The contributions from signal \bztodslv, resonant
and combinatorial backgrounds are indicated.}

By integrating the differential partial decay width (equation~\ref{e:decayw})
over all values of $\omega$, the branching ratio \brbtodslv\
was also determined to be
\[
\brbtodslv = ( \bdbrval \pm \bdbrstat )\,\%\, ,
\]
where again the error is statistical only. This result is consistent with
the world average once systematic errors are included.

Many previous results have been obtained using a constrained quadratic 
expansion for the form factor:
$\fw=\fone \left[1-a^2(\omega-1)+b(\omega-1)^2\right]$,
where $a$ is a slope parameter to be determined by the fit, and $b$ is
constrained to $b=0.66a^2-0.11$ \cite{fquad}. To allow comparison with
such measurements, the fit was also performed with this parameterisation
of \fw, giving the results
$\fvcb=(\fvcboval \pm \fvcbostat)\times 10^{-3}$ and
$a^2=\rhsqoval \pm \rhsqostat$,
the correlation between \fvcb\ and $a^2$ being \fvrhocrl. 
The difference in the two curvature parameters \rhsq\ and $a^2$ is in good 
agreement with the expectation of $\rhsq-a^2\approx 0.21$ \cite{clnff}.

\section{Systematic Errors}\label{s:syst}

Systematic errors arise from the uncertainties in the fit input parameters
given in Table~\ref{t:input}, the Monte Carlo modelling of the
signal $\omega$ resolution, the recoil
spectrum of $\rm b\rightarrow\ddstar\ell\bar\nu$ decays and selection
efficiencies, and possible biases in the fitting method.
The resulting systematic errors on the values of \fvcb, \rhsq\ and 
\brbtodslv\ are summarized in Table~\ref{t:syst} and described in more 
detail below.

\begin{table}[tp]
\centering

\begin{tabular}{l|c|c|c}
Error Source & $\Delta(\fvcb)/(\fvcb)$ (\%) & $\Delta\rhsq$ & 
$\rm\Delta Br/Br$ (\%) \\ \hline
\rb\ value                                & 0.2 & - & 0.4 \\
\bzero\ lifetime                          & 1.3 & - & 2.6 \\
\fbd\ value                  & $^{+2.8}_{-2.3}$ & - & $^{+5.5}_{-4.5}$ \\
\bratio{\dstar}{\dzero\pi^+}              & 1.0 & - & 2.1 \\
\bratio{\rm b}{\dstar\tau^-\bar{\nu}}         & 0.1 & 0.00 & 0.4 \\
\bratio{\rm b}{\dstar\rm D_s^{(*)-}}          & 0.3 & 0.01 & 1.3 \\
\bratio{\rm b}{\dstar X}                      & 0.0 & 0.00 & 0.1 \\
\bratio{\rm c}{\dstar X}                      & 0.0 & 0.00 & 0.0 \\
$\rm b\rightarrow\ddstar\ell\bar{\nu}X$ rate  & 0.6 & 0.05 & 2.9 \\
$\rm b\rightarrow\ddstar\ell\bar{\nu}X$ model & 4.1 & 0.19 & 3.5 \\
b fragmentation                           & 1.1 & 0.12 & 3.2 \\
\dzero\ decay multiplicity                & 1.2 & 0.08 & 2.2 \\
$\omega$ reconstruction                   & 2.2 & 0.12 & 2.1 \\
Lepton identification efficiency          & 1.2 &  -   & 2.3 \\
Vertex tag efficiency                     & 1.7 &  -   & 3.4 \\
Track reconstruction efficiency           & 0.1 & 0.10 & 4.3 \\
Tracking resolution                       & 2.4 & 0.04 & 4.4 \\
Fitting method                            & 1.9 & 0.05 & 0.6 \\
\hline
Total & \fvcbpsyst & \rhsqsyst & \bdbrpsyst  
\end{tabular}
\caption{\label{t:syst}
Summary of systematic errors on the measured values of
\fvcb, \rhsq\ and \brbtodslv\ for the inclusive analysis. The fractional
errors on \fvcb\ and \brbtodslv\ are given, whereas the errors
on \rhsq\ are absolute.}
\end{table}

\begin{description}
\item[Input quantities:] The various numerical fit inputs were each
varied according to the errors given in Table~\ref{t:input} 
and the fit repeated to assess the resulting uncertainties.

\item[$\bf\boldmath\rm b\rightarrow\ddstar\ell\bar\nu$ decays:] The calculation of
Leibovich~\etal \cite{ligeti} was used to simulate the recoil spectrum
of \btodshlv\ decays, assumed to be produced via the semileptonic decay
$\bzero\rightarrow\ddstar\ell\bar\nu$. 
Here \ddstar\ represents a P-wave orbitally
excited charm meson. The calculation predicts the recoil spectra and relative
rates of semileptonic decays involving both the narrow $\rm D_1$ and 
$\rm D_2^*$ states and  wide $\rm D_0^*$ and $\rm D_1^*$ states. All these
decays are suppressed close to $\omega=1$ by an extra factor of $(\omega^2-1)$
when compared with the signal \bztodslv\ decays. This reduces the 
uncertainty due to the rate of \btodshlv\ decays in the extrapolation of
the signal recoil spectrum to $\omega=1$. Non-resonant \btodshlv\ decays
are not included in the model, but are not expected to contribute close
to $\omega=1$.

The differential decay rates 
in \cite{ligeti} are given in terms of five possible 
expansion schemes and several unknown parameters: a kinetic energy term
\etake\ and the slopes of the Isgur-Wise functions for the narrow and wide
\ddstar\ states \tauh\ and \zetah. These parameters were varied within
the allowed ranges $-0.75<\etake<0.75\,$GeV, $-2<\tauh<-1$ and $-2<\zetah<0$
subject to the constraint that the ratio 
$R=\Gamma(\bar{B}\rightarrow D_2^*\ell\bar\nu)/
(\Gamma(\bar{B}\rightarrow D_1\ell\bar\nu)$ lie within the measured range
$R=0.37\pm 0.16$ \cite{alephdss,cleodss}. This excludes the expansion schemes
$A_\infty$ and $B_\infty$ of \cite{ligeti} and constrains the allowable
values of $\etake$ in the others. The fraction of \btodsslv\ decays involving
the narrow $\rm D_1$ and $\rm D_2^*$ states, which is not precisely
predicted, was varied in the range $0.22\pm 0.06$, obtained by comparing
the measured rates for \bzero\ semileptonic decays involving $\rm D^+$, \dstar,
$\rm D_1$ and $\rm D_2^*$ with the inclusive semileptonic decay rate 
\cite{pdg98,alephdss,cleodss}.
The systematic errors were determined as half the difference between the 
two parameter sets giving the most extreme variations in \fvcb\ and \rhsq,
and the central values were adjusted to  half way between these two extremes.
The values of both \fvcb\ and \rhsq\ are most sensitive to variations in
\etake, which is constrained by the measured value of $R$.

\item[b fragmentation:] The effect of uncertainties in the average 
b hadron energy $\meanxe=E_{\rm b}/E_{\rm beam}$ was assessed in Monte Carlo
simulation
by reweighting the events so as to vary \meanxe\ in the range $0.702\pm 0.008$ 
\cite{hfew}, and repeating the fit. The largest of the variations observed
using the fragmentation functions of Peterson, Collins and Spiller,
Kartvelishvili and the Lund group \cite{fragall}
were taken as systematic errors.

\item[$\bf\boldmath\dzero$ decay multiplicities:] 
The signal reconstruction efficiency and vertex charge distributions
are sensitive to the \bzero\ decay multiplicity, which depends only 
on the \dzero\ decay for the \bzero\ decay channels of interest.
The systematic error was assessed by varying
separately the \dzero\ charged and $\pi^0$ decay multiplicities in Monte Carlo
simulation
according to the measurements of Mark III \cite{markIII}. The branching
ratio $\dzero\rightarrow\rm K^0,\bar{K}^0$ was also varied according to its
uncertainty \cite{pdg98}. The resulting uncertainties on \fvcb\ and \rhsq\ 
from each variation were added in quadrature to determine the total
systematic errors.

\item[$\boldmath\omega$ reconstruction:] The modelling of the $\omega$ 
resolution depends on the description of the \dstar\ and \bzero\ energy
and angular distributions in the simulation.
The reconstructed \dstar\ and \bzero\
energy distributions in data and simulation were compared, and the means
were found to differ by 0.04 and 0.13\,GeV respectively.
 The opposite hemisphere missing energy was found to agree within
5\,\%. The corresponding systematics were assessed by shifting or scaling
the data distributions and repeating the $\omega$ reconstruction and fit.
The modelling of the angular resolution was checked by studying the agreement
of the two angular estimators---the slow pion and \dzero\ directions
for the the \dstar, and the missing energy vector and vertex flight directions
for the \bzero. The angular resolutions were found to be up to 5\,\% worse
in data, and the systematic error was assessed by degrading the simulated
resolution appropriately. Finally, the fraction of events with $\omega$
reconstructed in the physical region ($\omega<\omega_{\rm max}$) was found
to be 3.5\,\% smaller in the data, in both opposite and same sign charge 
samples. The reconstruction efficiency was corrected for this effect, and an
additional systematic error of half the correction ($1.7\,\%$) assumed. 
The final systematic
errors due to $\omega$ resolution modelling are dominated by the
\bzero\ $\theta$ resolution.

\item[Lepton identification efficiency:] The electron identification efficiency
has been studied using control samples of pure electrons from 
$\rm e^+e^-\rightarrow e^+e^-$ events and photon conversions, and found to
be modelled to a a precision of 4.1\,\% \cite{opalrb}.
The muon identification efficiency
has been studied using muon pairs producted in two-photon  collisions and
$\zb\rightarrow\mu^+\mu^-$ events, giving an uncertainty of 2.1\,\% 
\cite{opalbtol}.

\item[Vertex tag efficiency:] The fraction of hemispheres with identified
leptons which also had a selected secondary vertex was found to be about 
4\,\% less in data than in simulation. The overall fraction of
vertex tagged hemispheres was also found to be about 4\,\% lower in data.
These discrepancies were translated into systematic errors on the
efficiency to tag a semileptonic b decay with a secondary vertex in either
the same or the opposite hemisphere, in each case attributing the whole
discrepancy to a mismodelling of b hadron decays. The resulting errors on
the same and opposite hemisphere tagging efficiencies were taken to be fully
correlated.

\item[Track reconstruction efficiency:] The overall track reconstruction
efficiency is known to be modelled to a precision of 1\,\% \cite{opalrb},
and a similar uncertainty was found to be appropriate for the particular
class of slow pion tracks from \dstar\ decays. The systematic error was
assessed by randomly removing 1\,\% of tracks in the simulation and
repeating the fit.

\item[Tracking resolution:] Uncertainties in the tracking
detector resolution affect the efficiency, $\omega$ reconstruction and
vertex charge distributions. The associated error
was assessed in the simulation by applying a global
10\,\% degradation to all tracks, independently in the $r$-$\phi$ and
$r$-$z$ planes, as in \cite{opalrb}.

\item[Fit method:] The entire fitting procedure was tested on a fully
simulated Monte Carlo sample seven times bigger than the data, with true
values of $\fvcb=32.5\times 10^{-3}$ and $\rhsq=1.3$. The fit gave the results
$\fvcb=(31.8\pm 0.5)\times 10^{-3}$ and $\rhsq=1.25\pm 0.07$, consistent
with the true values. For each variable, the larger of the deviation of 
the result from the true value and the statistical error were taken as 
a systematic errors due to possible biases in the fit.
Additionally, the large Monte Carlo sample was reweighted to change
the values of \fvcb\ and \rhsq, and the fit correctly recovered the modified
values. To verify the correctness of the statistical errors returned by the 
fit, it was performed on many separate
subsamples, and the distribution of fitted results studied.
Further checks on the data included performing the analysis separately for 
\bzero\ decays involving electrons and muons, dividing the sample
according to the year of data taking, and varying the lepton 
transverse momentum cut. In all cases, consistent results were
obtained.

\end{description}
Including all systematic uncertaintities, the final result of the inclusive
analysis is
\begin{eqnarray*}
\fvcb & = & (\fvcbval \pm \fvcbstat \pm \fvcbsyst)\times 10^{-3}, \\
\rhsq & = & \rhsqval \pm \rhsqstat \pm \rhsqsyst,
\end{eqnarray*}
where the first error is statistical and the second systematic in each case.

\section{Measurement using exclusively reconstructed \boldmath\dstar\
decays}\label{s:excl}

In this analysis, the \dzero\ from the \bztodslv, 
$\dstar\rightarrow\dzero\pi^+$ decay is reconstructed exclusively in the
decay modes $\dzero\rightarrow\rm K^-\pi^+$ (`3-prong') and 
$\dzero\rightarrow\rm K^-\pi^+\pi^0$ (`satellite'---where the $\pi^0$ is
not reconstructed).
The event selection, reconstruction
and determination of $\omega$ are exactly the same as described in 
\cite{opalvcb}. The determination of the signal and background fractions,
and the fit to extract \fvcb\ and \rhsq\ are performed as in \cite{opalvcb}, 
but have been updated using the
improved form factor parameterisation \cite{clnff}, the updated 
input parameters given in  Table~\ref{t:input} and the 
$\rm b\rightarrow \ddstar\ell\bar\nu$ decay model \cite{ligeti} discussed
in Section~\ref{s:syst}.

The fit is performed on 814 3-prong and 1396 satellite candidates, of which
$505\pm 44$ and $754\pm 72$ are attributed to \bztodslv\ signal decays. The
result of the fit is
\begin{eqnarray*}
\fvcb & = & (\fvcbeval \pm \fvcbestat \pm \fvcbesyst)\times 10^{-3}, \\
\rhsq & = & \rhsqeval \pm \rhsqestat \pm \rhsqesyst\,,
\end{eqnarray*}
where again the first errors are statistical and the second systematic. 
The statistical correlation between
\fvcb\ and \rhsq\ is 0.95. The distribution of reconstructed $\omega$ 
for selected candidates (both 3-prong and satellite) is shown in 
Figure~\ref{f:exclw}. The branching ratio \brbtodslv\ has also been determined
to be
\[
\brbtodslv = (\bdbreval \pm \bdbrestat \pm \bdbresyst)\,\% .
\]

\epostfig{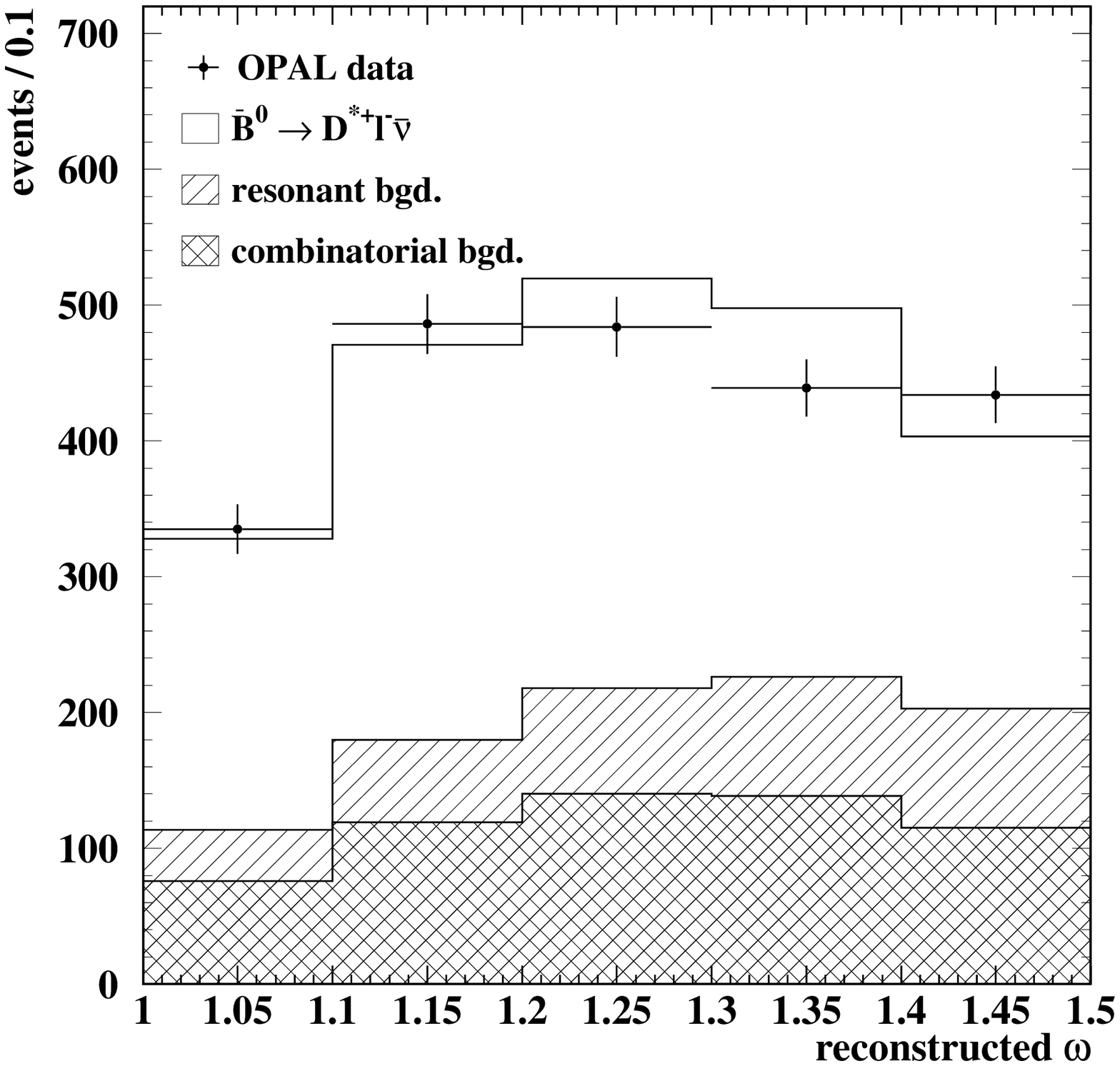}{f:exclw}{Distribution of reconstructed $\omega$ for
\bztodslv\ candidates in the exclusive analysis. The data are shown by the 
points with statistical error bars, and the fit result by the histogram. 
The contributions from signal \bztodslv, resonant and combinatorial backgrounds
are indicated.}

The systematic errors arise from uncertainties in the background levels in the
selected samples, as well as uncertainties in the Monte Carlo simulations.
They have been evaluated following the procedures described in \cite{opalvcb},
and are summarised in Table~\ref{t:syst2}. The selection efficiency error
includes contributions from lepton identification efficiency, b fragmentation
and detector resolution uncertanties, described in detail in \cite{opaldbr}.
The largest change with respect to the previous result is due to the
improved $\rm b\rightarrow\ddstar\ell\bar\nu$ modelling, with the suppression
of this background at values of $\omega$ close to one. This reduces the
statistical error, the systematic uncertainty due to the rate of such decays, 
and shifts the central value of \fvcb\ upwards as compared to \cite{opalvcb}.

\begin{table}[tp]
\centering

\begin{tabular}{l|c|c|c}
Error Source & $\Delta(\fvcb)/(\fvcb)$ (\%) & $\Delta\rhsq$ & 
$\rm\Delta Br/Br$ (\%) \\ \hline
\rb\ value                                & 0.2 & -     & 0.4 \\
\bzero\ lifetime                          & 1.2 & -     & 2.4   \\
\fbd\ value                               & 2.5 & -     & 4.9   \\
\bratio{\dstar}{\dzero\pi^+}              & 1.0 & -     & 2.0 \\
\bratio{\dzero}{\rm K^-\pi^+}             & 0.4 & 0.011 & 0.2  \\
\bratio{\dzero}{\rm K^-\pi^+\pi^0}        & 2.6 & 0.025 & 3.8  \\
\bratio{\rm b}{\dstar\tau^-\bar{\nu}}         & 0.1 & -     & 0.2 \\
\bratio{\rm b}{\dstar\rm D_s^{(*)-}}          & 0.2 & -     & 0.4 \\
$\rm b\rightarrow\ddstar\ell\bar{\nu}X$ rate  & 1.1 & 0.078 & 2.0 \\
$\rm b\rightarrow\ddstar\ell\bar{\nu}X$ model & 1.0 & 0.128 & 5.0 \\
Combinatorial background                  & 1.2 & 0.012 & 1.7  \\
Fake leptons                              & 0.2 &  -    & 0.4  \\
Fake \dzero\                              & 0.5 & 0.005 & 0.7  \\
$\omega$ reconstruction                   & 1.4 & 0.035 & -  \\
Selection efficiency                      & 2.9 & 0.005 & 3.1 \\
\hline
Total & \fvcbpesyst & \rhsqevsyst & \bdbrpesyst  
\end{tabular}
\caption{\label{t:syst2}
Summary of systematic errors on the measured values of
\fvcb, \rhsq\ and \brbtodslv\ for the exclusive analysis. The fractional
errors on \fvcb\ and \brbtodslv\ are given, whereas the errors
on \rhsq\ are absolute.}
\end{table}

\section{Conclusions}\label{s:conc}

The CKM matrix element \mvcb\ has been measured by studying the rate of the
semileptonic decay \bztodslv\ as a function of the recoil kinematics
of both inclusively and exclusively reconstructed \dstar\ mesons. The two
results are combined, taking into account the statistical correlation
of 18\,\% and correlated systematic errors from physics inputs and
detector resolution. The results are:
\begin{eqnarray*}
\fvcb & = & (\fvcbcval \pm \fvcbcstat \pm \fvcbcsyst)\times 10^{-3}, \\
\rhsq & = & \rhsqcval \pm \rhsqcstat \pm \rhsqcsyst\,, \\
\brbtodslv & = & (\bdbrcval \pm \bdbrcstat \pm \bdbrcsyst)\,\%\,,
\end{eqnarray*}
where the first result is statistical and the second systematic in each case.
The statistical and systematic correlations between \fvcb\ and \rhsq\ are
\statcorl\ and \systcorl\ respectively.
These results supersede our previous publication
\cite{opalvcb}. They are consistent with other determinations of
\fvcb\ at LEP \cite{alephvcb,delvcb} and  the $\Upsilon(4S)$ resonance 
\cite{upsvcb}. The branching ratio is consistent with the world average 
result of $(4.60\pm 0.27)$\,\% \cite{pdg98}.
The result for \fvcb\ is the most precise
to date from any single experiment.

Using the theoretical estimate $\fone=\foneval\pm\foneerr$ \cite{babar}, 
the value of \mvcb\ is determined to be
\[
\mvcb = (\vcbval \pm \vcbstat \pm \vcbsyst \pm \vcbtheo)\times 10^{-3},
\]
where the uncertainties are statistical, systematic and theoretical 
respectively.

\section*{Acknowledgements}

We particularly wish to thank the SL Division for the efficient operation
of the LEP accelerator at all energies
 and for their continuing close cooperation with
our experimental group.  We thank our colleagues from CEA, DAPNIA/SPP,
CE-Saclay for their efforts over the years on the time-of-flight and trigger
systems which we continue to use.  In addition to the support staff at our own
institutions we are pleased to acknowledge the  \\
Department of Energy, USA, \\
National Science Foundation, USA, \\
Particle Physics and Astronomy Research Council, UK, \\
Natural Sciences and Engineering Research Council, Canada, \\
Israel Science Foundation, administered by the Israel
Academy of Science and Humanities, \\
Minerva Gesellschaft, \\
Benoziyo Center for High Energy Physics,\\
Japanese Ministry of Education, Science and Culture (the
Monbusho) and a grant under the Monbusho International
Science Research Program,\\
Japanese Society for the Promotion of Science (JSPS),\\
German Israeli Bi-national Science Foundation (GIF), \\
Bundesministerium f\"ur Bildung, Wissenschaft,
Forschung und Technologie, Germany, \\
National Research Council of Canada, \\
Research Corporation, USA,\\
Hungarian Foundation for Scientific Research, OTKA T-029328, 
T023793 and OTKA F-023259.\\

\end{document}